\documentclass{jfm}

\usepackage[english]{babel}
\usepackage[usenames, dvipsnames]{color}
\usepackage[utf8]{inputenc}
\usepackage{graphicx}
\usepackage{caption}
\usepackage{subcaption}
\usepackage{natbib}
\usepackage{fixltx2e}
\usepackage{textgreek}
\usepackage{textcomp}
\usepackage[colorlinks=true,citecolor=blue,linkcolor=blue]{hyperref}
\usepackage{multirow}
\usepackage{array}
\usepackage{pbox}
\usepackage{siunitx}
\usepackage{xcolor}
\definecolor{cinnamon}{rgb}{0.82, 0.41, 0.12}
\definecolor{pink}{rgb}{0.858, 0.188, 0.478}
\definecolor{black}{rgb}{0.0, 0.0, 0.0}

\usepackage{amsmath,amssymb}

\begin{document}

\newtheorem{lemma}{Lemma}
\newtheorem{corollary}{Corollary}

\shorttitle{Effective interfacial tension in a flow-focusing channel} 
\shortauthor{K. Gowda, C. Brouzet, T. Lefranc, L. D. S\"{o}derberg and F. Lundell} 

\title{Effective interfacial tension in flow-focusing of colloidal dispersions: 3D numerical simulations and experiments}

\author{Krishne Gowda V. \aff{1}, Christophe Brouzet \aff{1,2}, Thibault Lefranc \aff{3}, 
	L.~Daniel S\"{o}derberg \aff{1,2}
	\and Fredrik Lundell\aff{1,2} \corresp{\email{fredrik@mech.kth.se}}}

\affiliation{\aff{1} Linn\'e FLOW Centre, KTH Mechanics, KTH Royal Institute of Technology, Stockholm SE-100 44, Sweden
	
	\aff{2} Wallenberg Wood Science Center, KTH Royal Institute of Technology, Stockholm SE-100 44, Sweden
	
	\aff{3} Univ Lyon, ENS de Lyon, Univ Claude Bernard, CNRS, Laboratoire de Physique, 
F-69342 Lyon, France }

\maketitle

\begin{abstract}

  An interface between two miscible fluids is transient, existing as a non-equilibrium state before complete molecular mixing is reached. However, during the existence of such an interface, which typically be at short timescales, composition gradients at the boundary between the two liquids cause stresses effectively mimicking an interfacial tension. Here, we combine numerical modelling and experiments to study the influence of an effective interfacial tension between a colloidal fibre dispersion and its own solvent on the flow in a microfluidic system. In a flow-focusing channel, the dispersion is injected as core flow that is hydrodynamically focused by its solvent as sheath flows. This leads to the formation of a long fluid thread, which is characterised in 3D using Optical Coherence Tomography and simulated using a volume of fluid method. The simulated flow and thread geometries very closely reproduce the experimental results in terms of thread topology and velocity flow fields. By varying the effective interfacial tension numerically, we show that it clearly influences threading dynamics and that it can be described by an effective capillary number. Furthermore, we demonstrate that the applied methodology provide means to measure the ultra-low but dynamically highly significant effective interfacial tension.

\end{abstract}
 
\section{Introduction}\label{sec:Introduction}
 Microfluidic techniques have facilitated the development of both fundamental and applied research in the field of chemistry, biology, medicine, material and physical sciences~\citep{squires2005microfluidics,Whitesides_2006, Stone_2004, Nunes_2013}. In particular, the flow-focusing configuration (illustrated in figure~\ref{fig1}), where a core fluid is focused by two sheath flows has been successfully employed in a wide variety of applications such as bubble or droplet formation~\citep{Cubaud_2008, Anna_2003,Garstecki_2004}, micro and nano-particle production~\citep{Mart_n_Banderas_2005}, hydrodynamic assembly of nanoparticle dispersion into high-performance superstructures~\citep{haakansson2014hydrodynamic,Ekanem_2015,Kamada:2017, Mittal_2018}, cell patterning~\citep{Takayama_1999}, DNA stretching~\citep{Wong_2003}, and diffusion-mixers~\citep{Knight:1998,Pollack_1999}. At the micrometre scale, phenomena such as interfacial tension and viscosity usually become dominant compared to the effects of gravity and inertia, which often are negligible. This accords unique characteristics to utilize multiphase flows in microfluidic devices and offers the possibility to thoroughly investigate the influence of fluid physiochemical properties on the flow.

For a pair of \textit{immiscible} fluids, surface or interface properties are controlled by interfacial tension. The relevant dimensionless number controlling the ratio between viscous forces and interfacial tension is the capillary number
\begin{equation}
Ca = \frac{\eta U}{\gamma},
\end{equation}
where $U$ is the typical velocity of the flow, $\eta$ the kinematic viscosity, and $\gamma$ the interfacial tension between the two fluids. The capillary number acts as the most important dimensionless number for dynamics of bubbles or droplets in microfluidics~\citep{Thorsen_2001, Anna_2003}. In the case of flow-focusing,
the capillary number is the primary dimensionless number controlling the transition between different flow patterns denoted threading, tubing, dripping and jetting~\citep{Cubaud_2008}. At very low capillary numbers, in the dripping regime, elongated droplets of length larger than the channel width are formed.  At large capillary numbers, the flow is in the threading regime where there is a formation of a continuous fluid thread. For intermediate capillary numbers, the thread can break up to form dripping and jetting flow patterns depending on the capillary instability mechanisms involved~\citep{humphry2009, utada2007dripping}. These flow patterns demonstrate the rich dynamics of flow features typically observed in a microfluidic device. Moreover, much of the effort in the last two decades has been devoted to understand the influence of system parameters like flow rates of two fluid phases, device geometry, surface chemistry, and material parameters, such as their viscosities, densities and interfacial tension, on the formation and motion of these flow patterns~\citep{Nunes_2013,anna2016}.

When a pair of \textit{miscible} fluids are in contact, there is a transient interface whose thickness is governed by the diffusion coefficient $D$.  Indeed, as microfluidic flows are mainly laminar, the fluid streams remain parallel and the liquids mix only by diffusion. One can then define the diffusion time scale $T_d=h^{2}/D$ and the convective time scale $T_c=h/U$, where $h$ is the characteristic length scale of the system. These time scales can be compared using the dimensionless Péclet number
\begin{equation}
Pe = \frac{T_d}{T_c} = \frac{hU}{D},\label{eq:peclet}
\end{equation}
which is the relevant dimensionless number to describe convective diffusive flows~\citep{Cubaud:2006,atencia2004}. When miscible fluids are subjected to flow-focusing, at high Péclet number, the main flow regime observed is a threading regime with common features similar to immiscible fluid threads while, at low Péclet number, the thread can be destabilized by diffusion instabilities~\citep{CubaudNotaro2014}. However, as the existence of such thread structures is transient and limited to short timescales before complete mixing, little is still known about the dynamics of viscous threads in miscible environments.

From the above, it could be expected that the dynamics of immiscible and miscible fluids in microfluidic environments are controlled by the capillary and Péclet numbers, respectively. However, miscible fluids flowing under laminar flow conditions and high Péclet number have a boundary created by so called \textit{de facto} dynamic interface, which gradually changes with time~\citep{joseph1990fluid}. Such an interface may resemble a distinct interface present between two immiscible fluids~\citep{garik1991interfacial,anderson1998diffuse}. The \textit{de facto} dynamic interface arises due to non-equilibrium capillary forces generated by the interfacial stresses whenever gradients of a fluid property exist at the interface of two miscible fluids~\citep{Korteweg1901}. Recently,~\cite{truzzolillo2014} coined the term effective interfacial tension (EIT) for this effect, and defined it as \begin{equation} \gamma = K \frac{\Delta\Phi^{2}}{\delta},\end{equation}
where $K$ is the Korteweg constant of the system, $\delta$ is the thickness of the interface and $\Delta \Phi$ is the change in the concentration or volume fraction $\Phi$. They noted that despite Korteweg's theory being a century old, experimental investigations are limited to molecular fluids. 
They carried out experimental measurements of the Saffman-Taylor instability in a Hele-Shaw cell with miscible complex fluids including colloidal dispersions. Their measurements went beyond confirming Korteweg's theory and showed that EIT is also a function of particle structure (basically size and shape) together with  particle-particle and particle-solvent interactions~\citep{truzzolillo2016}.

Effective interfacial tension is expected to have relevance in many fields starting from material processing to multiphase complex fluid dynamic problems involving droplet and bubbles formation, jetting, coalescence and break-up of droplets~\citep{truzzolillo2017}. Even though  progress has been made on the understanding of the above aspects, transient mechanisms and their effect on the formation of the above flow patterns are still unexplored. To the best of our knowledge, there are no reports detailing the significance of EIT on the flow physics in microfluidic systems. Moreover, the recent review by \cite{anna2016} observed that many of the advancements in the field of microfluidics have been driven predominantly by experiments due to the complexity of geometries and fluid physics. Thus, there is a demand for numerical simulations that can provide insights on complex mechanisms, and to extract the variables that are difficult or impossible to measure in experiments. 

We aim at addressing the influence of EIT and its effects on the flow-focusing of a colloidal dispersion by its own solvent. Detailed 3D experimental and numerical investigations have been performed in a microfluidic flow-focusing channel of square cross-section using Optical Coherence Tomography and Volume of Fluid method.
The colloidal dispersion is formed by cellulose nanofibrils (CNF) dispersed in water. Such a colloidal fibre dispersion has been used to assemble very strong cellulose filaments through hydrodynamic focusing~\citep{haakansson2014hydrodynamic, Mittal_2018}. Furthermore, ~\cite{truzzolillo2017} noted that for such a system, EIT is expected to exist between the colloidal dispersion and its solvent.  Here, we quantitatively investigate the characteristics of thread formation such as 3D topology and velocity fields as a function of EIT and flow-rate ratio. The previous experimental studies by \cite{Cubaud_2008, cubaud2009} and~\cite{CubaudNotaro2014} for both immiscible and miscible high-viscosity threads in flow-focusing channels  were limited to a top view and did not resolve the full 3D shape of the thread. Indeed, there are only a few 3D studies of the threading regime~\citep{Knight:1998,bingrui2016} but their focus are not on the effect of interfacial properties on the thread dynamics.
Thus, at large Péclet number, the role of EIT potentially acting during the existence of a transient interface between two miscible fluids is not understood so far due to lack of 3D experimental and numerical studies.

This paper is organized as follows. In Section~\ref{methods}, experimental and numerical setups are presented together with the numerical validation. In Section~\ref{Threadshape}, we compare the results of 3D computations with the experimental measurements in the threading regime. In particular, the focus is on the evolution of 3D thread shape, and morphology of the region wetted by the colloidal dispersion on the top and bottom walls. The effects of flow-rate ratio on variation in thread shape is investigated. We also examine numerically the effect of EIT on the dynamics of thread formation, allowing us to estimate the value of EIT for the present colloidal dispersion system. In Section~\ref{velocity_fields}, we look at the velocity fields both along the centreline as well as in cross-sectional planes of the channel. Finally, in Section~\ref{conclusion}, we present the conclusions.

\section{Methodology}\label{methods}
\subsection{Experimental set-up}
\subsubsection{Flow cell}\label{flow_cell_fluids}

\begin{figure} 
		\centering
	\includegraphics[width=0.75\linewidth]{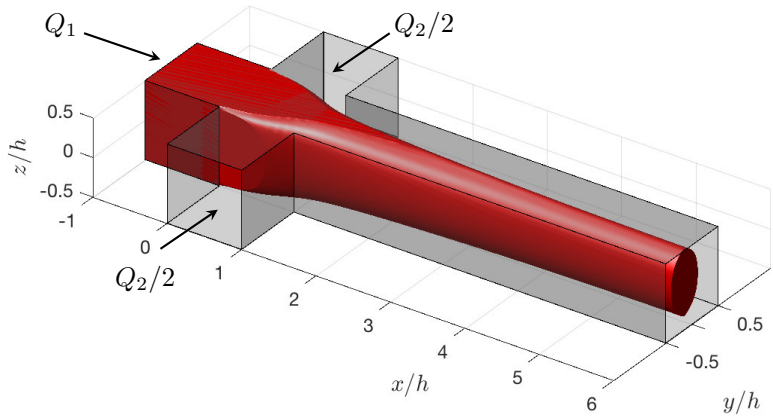}
	\caption{\small Schematic of the flow-focusing channel.} 
	\label{fig1}
\end{figure}

In the present work, we use a flow-focusing channel system with square cross-sections having  $h = 1$~mm sides as shown schematically in figure~\ref{fig1}. The channel system comprises four channels, with three inlet channels intersecting to form a cross-junction, and a main outlet channel for the fluids to flow out. A Cartesian coordinate system ($x,y,z$) is introduced, with the $x$ and $y$ axes in directions of the core and sheath flow inlets, respectively. The $z$ axis is perpendicular to the $(x,y)$-plane representing the third direction. Here, a colloidal dispersion in the core flow is focused by two water sheath flows. The flow-focusing channel is fabricated out of a $1$~mm thick stainless-steel plate~\citep{haakansson2014hydrodynamic}. The steel plate is covered on both sides with a $140~\mu$m Cyclic Olefin Copolymer (COC) film providing optical access. The COC - channel plate - COC sandwich is held together by two aluminium plates with screws. Two syringe pumps (WPI, AI-$4000$) were used to drive the core and sheath flows at constant volumetric flow rates. In most of the experiments, flow rates are $Q_{1}$ = 6.5  mm$^{3}\cdot$s$^{-1}$ for the colloidal dispersion and $Q_{2}$~=~ $7.5$~mm$^{3}\cdot$s$^{-1}$ for the sheath flows, respectively. The ratio of these flow rates is defined as $\phi=Q_1/Q_2$ and is set to $\phi=0.8667$ in the standard case, except when explicitly varied in Section~\ref{flow_rate_ratio}. The indices 1 and 2 will be used to denote the core and sheath fluid/flow respectively, throughout the paper.

\subsubsection{Colloidal dispersion }\label{fibre dispersion}
The colloidal dispersion used as the core fluid is cellulose nanofibrils (CNF) suspended in water. Cellulose nanofibrils were prepared by liberating fibrils from bleached softwood fibres (Domsjö Fabriker AB, Sweden). Before liberation, the fibrils were carboxymethylated~\citep{Wagberg1987163} to a degree of substitution of $0.1$. The fibrils were then liberated from the fibre wall following a protocol described in~\cite{fall2011colloidal}. Post-liberation unfibrillated fibre fragments were removed by centrifuging the dispersions at {4500} {rpm}. A transparent colloidal dispersion of concentration $3$~g$\cdot$dm$^{-3}$ was obtained with typical fibril length~$l$ of $700$~nm and fibril diameter~$d$ of $2$ to $3$~nm. 

\begin{figure}
	\centering{\includegraphics[width=0.75\textwidth]{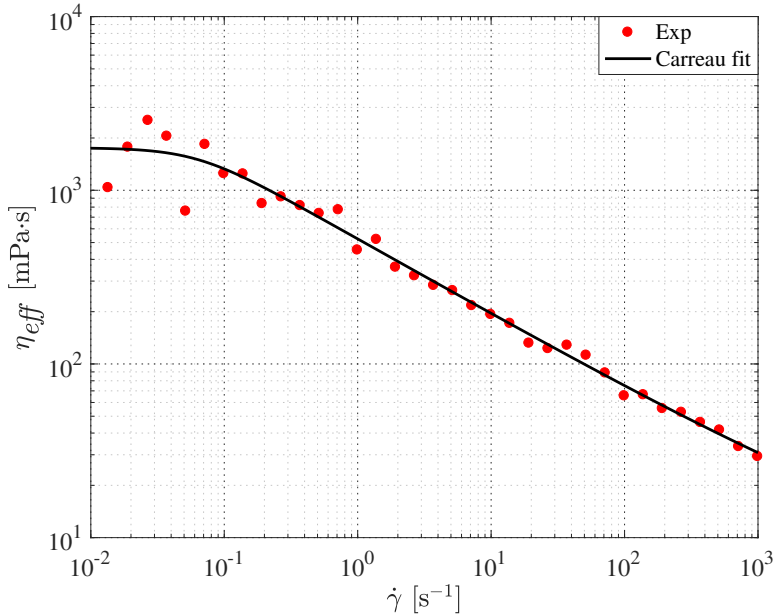}}
	\caption{ Shear viscosity measurements of the colloidal dispersion represented by red dots, and the non-Newtonian Carreau model fit (equation~\ref{Car}) indicated by the black solid line.}
	\label{fig2:Rheology_measurement}
\end{figure}

The rheological characterization of the colloidal dispersion was carried out using a Kinexus pro+ rheometer (Malvern) with a modified Couette geometry, suitable for accurate measurement of viscosity at low shear rates. The shear viscosity behaviour of the colloidal dispersion is indicated by red dots as shown in figure~\ref{fig2:Rheology_measurement}. It exhibits a classical shear thinning behaviour at moderate and high shear rates, and constant viscosity at low shear rates~\citep{Martoiaetal2016,Nechyporchuketal2016}. The rheological data can be fitted well by a Carreau model
\begin{equation}
\eta_{\textrm{\textit{eff}}}=\eta_{\textrm{\textit{inf}}}+(\eta_{0}-\eta_{\textrm{\textit{inf}}})[1+(\tau\dot{\gamma})^2]^\frac{n-1}{2},
\label{Car}
\end{equation}
where $\eta_{\textrm{\textit{eff}}}$ is the shear viscosity, $\dot{\gamma}$ the shear rate, $\eta_{0}$ the zero shear viscosity, $\eta_{\textrm{\textit{inf}}}$ the infinite shear viscosity, $\tau$ the relaxation time, and $\mathit{n}$ the power index. The best fit is shown by the black solid line in figure~\ref{fig2:Rheology_measurement} and gives $\eta_{\textrm{\textit{inf}}}=5$~mPa$\cdot$s, $\eta_0=1756$~mPa$\cdot$s, $\tau=16.16$~s, and $n=0.56$. The viscosity ratio between the core and sheath flow defined as $\chi = \eta_{1}/\eta_{2} = 1756 $  is arrived by considering the viscosity of colloidal dispersion at low shear rate $\eta_{1}$ = $\eta_0=1756$~mPa$\cdot$s and viscosity of sheath fluid $\eta_2=1$~mPa$\cdot$s, respectively.

\subsubsection{System time scales}\label{time scales}
Due to their dimensions, the fibrils are Brownian and diffuse in the solvent. The diffusion coefficient~$D$ is given by~\citep{Doi1986}
\begin{equation}
D = \frac{k_{B} T \ln(l/d)}{2\pi \eta l},
\end{equation}
with the temperature $T= 300$~K, the Boltzmann constant $k_B= 1.38 \times 10 ^{-23}$~J$\cdot$K$^{-1}$, and the solvent viscosity $\eta= 1 $~mPa$\cdot$s. Plugging in the all the values, {this leads to }$D \approx 5 \times 10 ^{-12}$~m$^{2}\cdot$s$^{-1}$ and to a typical diffusion time scale in the system $T_d \approx 2 \times 10 ^{5}$~s. With a velocity of the flow $U \approx 10 $~mm$\cdot$s$^{-1}$, the convective time scale can be estimated to be $T_{c} \approx 10^{-1}$~s. Using these values and equation~(\ref{eq:peclet}), we get  $Pe \approx 2 \times 10 ^{6}$. The high Péclet number implies that the system is predominantly convective driven and diffusion is very weak. Therefore, it is possible for a \textit{de facto} dynamic interface to exist between the two fluids~\citep{joseph1990fluid, Korteweg1901}. The properties of a \textit{de facto} dynamic interface may resemble a distinct interface between two immiscible fluids and thus, the two fluids, the colloidal dispersion and water, exhibit a transient immiscible behaviour.  
In the following two sections, the experimental method and numerical setup will be presented. Experimental method involves more about the description of optical coherence tomography (OCT) employed for topology and velocity measurements.

\subsubsection{Optical Coherence Tomography}

\begin{figure}
	\centering
	\includegraphics[width=0.47\linewidth]{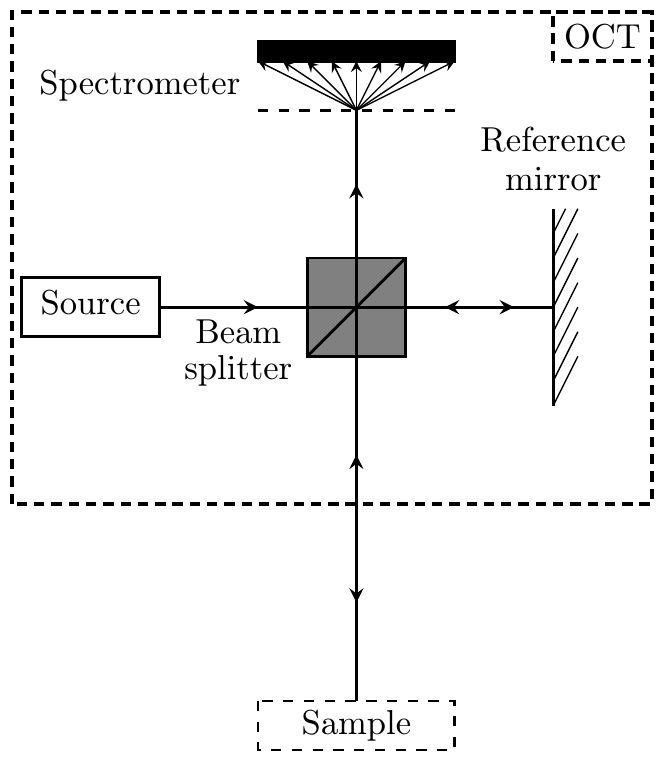}
	\caption{Schematic of the working principle of OCT, adapted from~\cite{tomlins2005theory}. \label{fig:fig3-working-principle-OCT}}
\end{figure}

\cite{huang1991optical} first demonstrated the three-dimensional non-invasive imaging technique known as Optical Coherence Tomography (OCT). Based on low-coherence interferometry, OCT is used to produce an image of optical scattering within transparent or turbid millimetric samples with a few micrometre resolutions. Usually designed for biomedical applications, the characteristics of the OCT, namely high-resolution, depth-resolved information, contact-free imaging and velocity acquisition, make it a pertinent measurement technique for a large range of other applications~\citep{stifter2007beyond}, e.g. material characterisation and microfluidics~\citep{xi2004structural,ahn2008optical}.

Since optical coherence tomography might be new to the reader, a brief description is in place. The interferometric technique used in OCT is based on the Michelson interferometer. Its principle is sketched in figure~\ref{fig:fig3-working-principle-OCT}. A broadband light source is divided into two parts by a beam splitter. One part is sent in a reference arm while the other goes through the sample. The reflected light from the sample is then recombined in the beam splitter with the light from the reference arm, providing an interference signal. Interferences are only observed when the reference and sample arm optical path lengths are matched to within the coherence length of the light~\citep{tomlins2005theory}. Therefore, the depth resolution of the system is determined by the coherence of the light source. A full interference pattern, i.e. interferences as a function of depth, can be obtained by analysing the output spectrum, thanks to the spectrometer. In the interference pattern, refractive index variations or differences of optical scattering between different layers in the sample correspond to intensity peaks or to different magnitudes of intensity. There, the interference pattern contains information about the depth and the optical scattering within the sample, with a micrometric resolution. Moreover, the depth and lateral resolutions are decoupled.

This imaging technique can be combined with a Doppler acquisition when there is a flow in the sample~\citep{Chenetal1997}. The backscattered light from moving particles experiences a double Doppler shift (one from the source, one from the particle back to the probe). Thereby, the velocity of moving particles can be measured by using the Doppler shift and the relative angle between the optical beam and the direction of the flow. Again, the depth resolution depends on the coherence length of the light source and is decoupled from the lateral resolution. The velocity resolution depends on the acquisition time and the scan angle. The range of velocities that can be measured is wide, varying from several micrometres per second to tens of millimetre per second.

The OCT apparatus used in this study is a commercial spectral domain OCT (SD-OCT) Telesto II from Thorlabs. The light source has a central wavelength of $1310$~nm and a bandwidth of $270$~nm, giving a resolution in both the axial and transverse directions of 3~$\mu$m. Here, OCT together with Doppler acquisition is employed for 3D measurements of thread topology and velocity fields. More details on the OCT acquisition and data processing can be found in Appendix~\ref{OCT}.

\subsection{Numerical set-up}
\subsubsection{Numerical method}
The numerical simulation of multiphase flows with fluid-fluid interfaces faces two major challenges. First, the accurate estimation and advection of the complex interface between two incompressible fluids. Second, suppression of spurious currents/velocities arising from inaccurate calculations of interface curvature. Thus, the demand for highly resolved simulations in terms of spatial resolution at the liquid-liquid interface has become pertinent to capture a sharp interface and analyse the physics of such flows. 

Volume of Fluid (VOF)~\citep{hirt1981volume} and level set methods\citep{osher1988fronts}, have received significant attention and are the most popular state-of-the-art tools in simulating multiphase flow problems involving extensive interface movement and topological changes. Here, we employ the VOF methodology, where the interface is implicitly represented using an indicator function, a fluid fraction parameter $\alpha$ \textit{via} the volume fractions of the two fluids in the computational cells; $\alpha$ = 0  correspond to fluid 1, $\alpha$ = 1 to fluid 2,  with 0 $<$ $\alpha$ $<$ 1  in the interface region. The advection of the interface is achieved through the redistribution of $\alpha$ between neighbouring cells. Various VOF methods have been developed since its inception, and are broadly categorised into algebraic and geometric methods. Algebraic VOF methods are developed for general mesh types, easier to implement and faster~\citep{deshpande2012evaluating}. However, they are less accurate and often require higher order schemes to retain the sharpness and boundedness of fluid fraction~$\alpha$. Geometric VOF methods, on the other hand, use geometric operations to reconstruct the interface inside a computational cell, and are more accurate, but also complex to implement and the execution is relatively slow.
In the present work, we have chosen a new geometric VOF approach called IsoAdvector~\citep{roenby2016computational} which combines high accuracy with advection of interface.

The full system of equations being solved consists of the Navier-Stokes equation
\begin{equation}
\frac{\partial
	{\rho_{b} \mbox{\boldmath $U$}}}{\partial t} +  \mbox{\boldmath $\nabla$}\cdot (\rho_{b}\mbox{\boldmath $U$}{\mbox{\boldmath $U$}} )= -{\mbox{\boldmath $\nabla$}} {p} + {\mbox{\boldmath $\nabla$}} \cdot {\mbox{\boldmath $T$}}+ \rho_{b}{\mbox{\boldmath $g$}} + {\mbox{\boldmath $F_s$}},
\label{momentum}
\end{equation}
the continuity equation
\begin{equation}
\mbox{\boldmath $\nabla$}\cdot \,
\mbox{\boldmath $U$} = 0,
\label{continuity}
\end{equation}
and the equation for the advection of fluid fraction $\alpha$
\begin{equation}
\frac{\partial \alpha}{\partial t} + \mbox{\boldmath $\nabla$}\cdot (\alpha\mbox{\boldmath $U$})  = 0.
\label{diffDim}
\end{equation}
Here $\textbf{\textit{U}}$ is the velocity vector field, $\rho_{b}$ the density, $p$ the pressure field, $\textbf{\textit{T}}$ the deviatoric stress tensor $( \mbox{\boldmath $T$} = 2 \mu_{b} \mbox{\boldmath $S$}    - 2 \mu_{b} (\mbox{\boldmath $\nabla$}\cdot \mbox{\boldmath $U$})  \textit{\textbf{I}} / 3 )$ , 
where \textit{\textit{\textbf{S}}}= $ 0.5 (\mbox{\boldmath $\nabla$} \mbox{\boldmath $U$} + \mbox{\boldmath $\nabla$} \mbox{\boldmath $U$}^ T )$, \textit{\textbf{I}} is the identity matrix and $\textbf{\textit{g}}$ is the gravitational force. The bulk parameters density $\rho_{b}$  and viscosity $\mu_{b}$ are computed based on the weighted average distribution of the fluid fraction $\alpha$

\begin{equation}
\rho_{b} = \rho_{1}\alpha + {\rho_{2}} (1 -\alpha),
\end{equation}

\begin{equation}
\mu_{b} = \mu_{1}\alpha + {\mu_{2}} (1 -\alpha),
\end{equation}

\noindent where $\rho_1$, $\rho_2$, $\mu_{1}$, ${\mu_{2}}$ are the densities and the viscosities of the two phases. ${\mbox{\boldmath $F_s$}}$ is the body force per unit mass, which accounts for the surface tension forces present only at the interface between two fluids. The surface tension force ${\mbox{\boldmath $F_s$}}$ is modelled as a volumetric force using Continuum Surface Force (CSF) method~\citep{Brackbill1992}, and is defined as

\begin{equation}
{\mbox{\boldmath $F_s$}=\gamma \kappa(\mbox{\boldmath $\nabla$} \alpha)}, 
\end{equation}
where $\gamma$ is the interfacial tension and $\kappa$ is the curvature of the interface

\begin{equation}
{\kappa = \mbox{\boldmath $\nabla$} \cdot \bigg(\frac{\mbox{\boldmath $\nabla$} \alpha}{|\mbox{\boldmath $\nabla$} \alpha|}\bigg)}.
\end{equation}
\\
The numerical simulations were performed with the finite-volume-based open-source code OpenFOAM~\citep{Weller_1998}. The PISO (pressure-implicit with splitting of operators)
scheme~\citep{Rusche_2002} is applied for solving the momentum balance equation~(\ref{momentum}) in conjunction with the continuity equation~(\ref{continuity}). More detailed information about the implementation of the above equations in the OpenFOAM framework can be found elsewhere~\citep{deshpande2012evaluating,Nekouei_2017}. A first order implicit Euler scheme was used for transient terms, and the time step was controlled by setting the maximum Courant number to 0.3. A second order Total Variation Diminishing (TVD) scheme with van Leer limiter was used for the spatial discretization. Boundedness of the fluid phase fraction $\alpha$, was controlled by  the IsoAdvector~\citep{roenby2016computational} scheme implemented in the \textit{interIsoFoam} solver of OpenFOAM-v1706~\citep{OpenCFD}.

Our system has three inlets, two for the sheath flows and one for the core flow, and one outlet as depicted in figure~\ref{fig1}. The length of three inlets and outlet channel correspond to $10$~mm and $30$~mm with square cross-section $1 \times 1$~mm$^{2}$, respectively. The non-Newtonian rheology of the core fluid has been implemented numerically using the Carreau model depicted in figure~\ref{fig2:Rheology_measurement}. We assume that the viscosity~$\eta_{\textrm{\textit{eff}}}$ is sufficient to describe the relevant material properties of the colloidal dispersion. To initialize the simulation, we set the fluid phase fraction, $\alpha = 0$ in the core flow inlet channel, $\alpha = 1$ in the sheath flow inlet channel and for the outlet channel. A uniform velocity was set at the inlets calculated as per the flow rates mentioned in Section~\ref{flow_cell_fluids}. At the walls, we apply the no-slip boundary condition and use an equilibrium contact angle $\theta = 0^{\circ}$ considering full wetting by the sheath fluid. The contact angle boundary condition was employed to correct the surface normal vector, and thereby adjusts the curvature of the interface in the vicinity of the wall. At the outlet, we prescribe a reference pressure (atmospheric) and zero gradient of the volume fraction 
\begin{equation}
\partial  \alpha/ \partial n = 0,
\end{equation}
where $n$ is the unit vector normal to the wall.
\subsubsection{Validation and convergence}
\begin{figure} 
	\centering
	\includegraphics[width=0.9\linewidth]{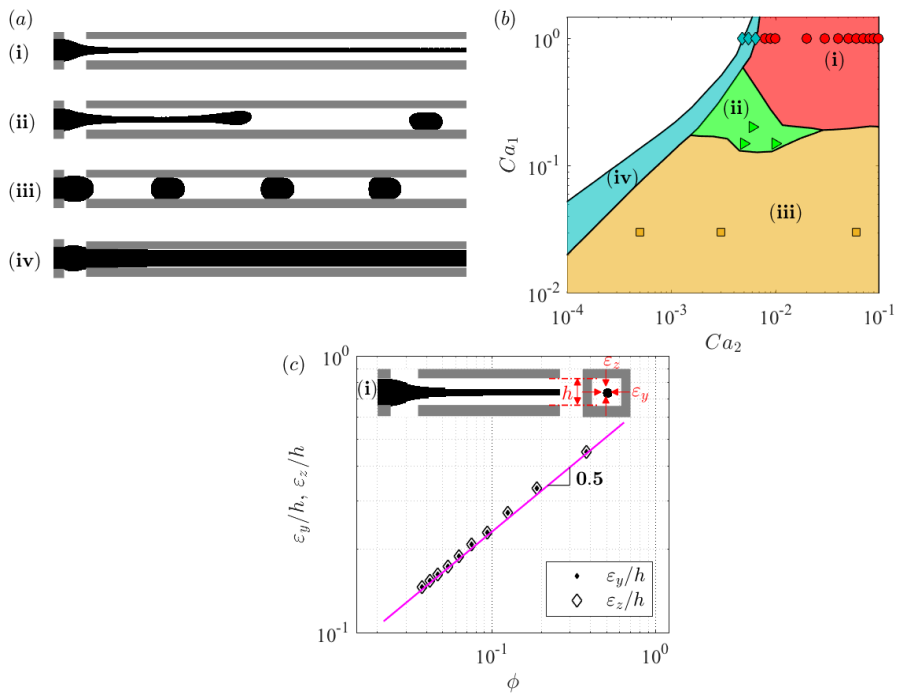}
	\caption{Validation of three-dimensional simulations in a microfluidic flow-focusing channel comparing with the observations of \cite{Cubaud_2008} experiments. (a)~Flow-patterns viz$.,$ (i) threading,(ii) jetting, (iii) dripping, and (iv) tubing; (b)~State space of flow regimes mapped based on capillary numbers of the core $Ca_{1}$ and sheath $Ca_{2}$ fluid phases, adapted from \cite{Cubaud_2008}. The markers show simulated cases: $\circ$ threading, $\triangleright$ jetting, $\square$ dripping and $\diamond$ tubing. (c)~Dimensionless width $\varepsilon_{y}/h$ and height $\varepsilon_{z}/h$ of the thread versus flow rate ratio~$\phi$. Inset: Top and cross-sectional view of flow-focusing channel depicting the flow-pattern of (i) threading regime, channel width $\mathit{h}$, width of thread $\varepsilon_{y}/h$ and height of thread $\varepsilon_{z}/h$.}
	\label{fig:fig4validation}
\end{figure}
The validation of the numerical scheme is performed by simulating a reference case, which also demonstrates the rich dynamics of flow in the flow-focusing geometry. The \cite{Cubaud_2008} experiments involve two immiscible Newtonian fluids subjected to hydrodynamic focusing achieved through a square microfluidic flow-focusing channel of side, with $h = 100~\mu$m. They have shown that the different regimes can be represented in a capillary number based flow map. Their experiments have been reproduced numerically here using glycerol and PDMS oil as core ($Q$$_{1}$) and sheath ($Q$$_{2}$) flows. These fluids have viscosities of $\eta_{1} = 1214$~mPa$\cdot$s, and $\eta_{2} = 4.59$~mPa$\cdot$s, respectively. The interfacial tension between the two fluids is $\gamma_{12} = 27.0$~mN$\cdot$m$^{-1}$. The 
capillary numbers of core and sheath flows are defined as $Ca_{1}$ = ${\eta_{1} Q_{1}}/{(\gamma_{12} h^{2})}$ and  $Ca_{2}$ = ${\eta_{2} Q_{2}}/{(\gamma_{12} h^{2})}$, respectively.
Figure~\ref{fig:fig4validation}~(a) shows the typical flow patterns viz., (i) threading, (ii) jetting, (iii) dripping, and (iv) tubing obtained through our 3D simulations. The state space in figure~\ref{fig:fig4validation} (b) shows that the agreement between flow regimes observed by \cite{Cubaud_2008} and in our simulations is very good. In particular, the transition between threading and tubing (from circles to diamonds) is well captured. In addition, the agreement with the experiments is further confirmed quantitatively by measuring the width $\varepsilon_{y}/h$ and height $\varepsilon_{z}/h$ of the thread in the (i) threading regime as a function of the flow rate-ratio~$\phi$ as depicted in figure~\ref{fig:fig4validation}~(c). Indeed, the thread cross-section is circular and scales as $\varepsilon_y/h=\varepsilon_z/h =~(\phi/2)^{1/2}$, observed experimentally and expected theoretically by~\cite{Cubaud_2008,cubaud2009}. The pink solid line represents the fit of $\varepsilon_y/h=\varepsilon_z/h = (\phi/2)^{1/2}$.
 \begin{figure} 
	\centering
	\includegraphics[width=0.95\linewidth]{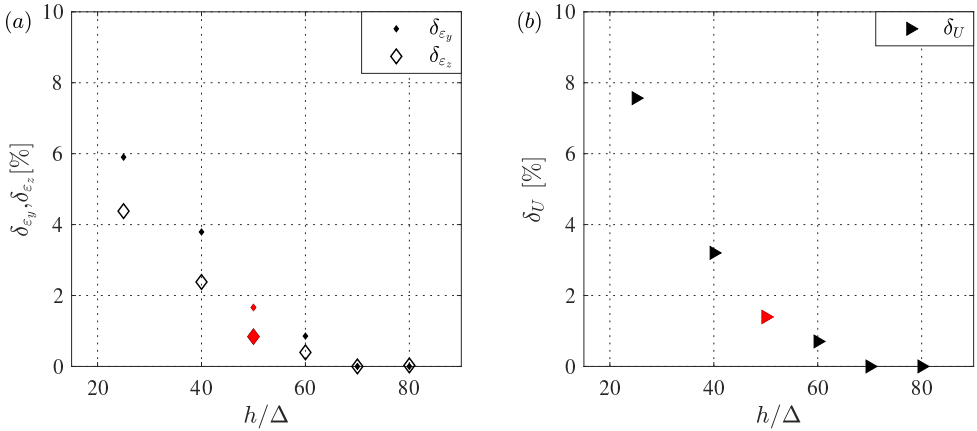}
	\caption{ Effect of grid size $\Delta$ on the (a) width $\varepsilon_{y}/h$, height $\varepsilon_{z}/h$ of the thread, and (b) velocity in a square channel cross-section of width $h$. The variation with respect to the finest resolution $h/\Delta=80$ in $\varepsilon_{y}/h$, $\varepsilon_{z}/h$, and velocity $\mathit{U}$ due to grid size $\Delta$ is indicated by $\delta_{\varepsilon_{y}}$, $\delta_{\varepsilon_{z}}$  and $\delta_{U}$. Red markers denote that a grid size $h/\Delta = 50$ which is chosen for the main results in the present study.}
	\label{fig:fig5gridstudy}
	\end{figure}

All the validation studies and upcoming results of the present study in Sections~\ref{Threadshape} and~\ref{velocity_fields} were obtained at a resolution of $h/\Delta = 50$, where $\Delta$ is the grid size, i.e., $50 \times 50$ cells in a square channel cross-section. Figure~\ref{fig:fig5gridstudy} shows the sensitivity of the grid size $\Delta$ on width $\varepsilon_{y}/h$, height $\varepsilon_{z}/h$  of the thread, and on the flow field velocity  $\mathit{U}$. We observe that for $h/\Delta = 50$, the variation with respect to the finest resolution $h/\Delta= 80$ in the width  ($\delta_{\varepsilon_{y}}$), height ($\delta_{\varepsilon_{z}}$) of the thread, and on the flow field velocity ($\delta_{U}$) is smaller than $2\%$, while for $h/\Delta> 60$ it is smaller than~$1\%$. It will be seen that $h/\Delta = 50$ is sufficient to capture the physics of flow with desired quality both in quantitative and qualitative forms. Typical clock time for the simulations range from $24$ to $48$~hours using $64$ to $256$~processors going from coarse to fine grid.

\section{Thread shape}\label{Threadshape}
In this section, we focus on the shape of the thread through a detailed comparison between experimental measurements and numerical simulations. In particular, the influence of flow-rate ratio $\phi$ and effective interfacial tension $\gamma$ on thread formation is investigated. 
The flow-rate~ratio and the effective interfacial tension (EIT) for the standard case are respectively set to $\phi = 0.8667$ and $\gamma~=~0.054$~mN$\cdot$m$^{-1}$ in all simulations, except when explicitly varied in Sections~\ref{flow_rate_ratio} and ~\ref{surface_tension}.
\begin{figure}
	\centering
	\includegraphics[width=1\linewidth]{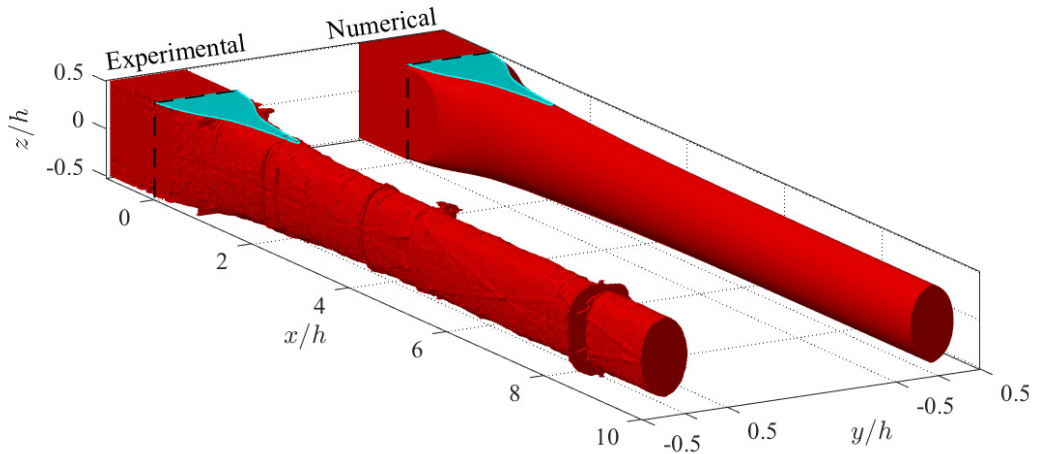}
	\caption{3D view of the experimental and numerical threads. The black dashed lines defines the beginning of the focusing region at $x/h = 0$,  and cyan colour at the top plane $z/h=0.5$ denotes the wetted region. 
	}
	\label{fig:fig6_Octthread}
\end{figure}

\subsection {Shape of the thread in the standard case\label{shape_thread}} 

\begin{figure} 
	\centering
	\vspace{0.5cm}
	\includegraphics[width=1\textwidth]{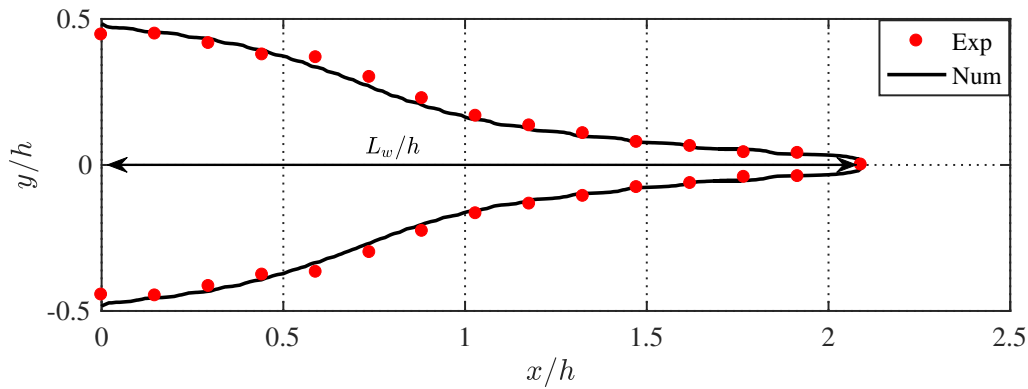} 
	\caption{ Comparison of experimental and numerical wetted regions of the core fluid in the plane $z/h=0.5$. The length of this region in the downstream direction~$x/h$ is defined as $L_w/h$. }
	\label{fig:fig7_Wetting_length}
\end{figure}

The 3D shapes of the thread region occupied by the colloidal dispersion measured using OCT and obtained by numerical simulations are shown in figure~\ref{fig:fig6_Octthread}. The qualitative agreement is very good (quantitative comparisons will follow later). For $x/h<0$, the 
colloidal dispersion attach to the channel walls. The beginning of focusing region where the entry of sheath flows commence corresponds to $x/h=0$. By inspecting figure~\ref{fig:fig6_Octthread} carefully at the position where the sheath flows are injected (0 $\leq$ $x/h$ $\leq$ 1), it can be noted that the colloidal dispersion does not lose contact with the walls located at $z/h =\pm 0.5$ immediately. Instead, there are regions wetted by the colloidal dispersion before detaching completely from the top and bottom walls as highlighted by cyan colour in figure~\ref{fig:fig6_Octthread}. Further downstream at around $x/h=2$, the cross-section of the colloidal dispersion thread forms a nearly elliptically shaped thread. The very good agreement between numerical and experimental results is again demonstrated in figure~\ref{fig:fig7_Wetting_length} where the experimentally and numerically obtained morphologies of the region wetted by the colloidal dispersion on the wall at $z/h=0.5$ are compared. The distance from $x/h=0$ to the point where the colloidal dispersion detaches from the wall is named wetted length and is denoted by $L_w/h$. All features such as wetted length $L_w/h$ and the morphology of the wetted region observed in experiments are very well captured and reproduced by the numerical simulations. 

\begin{figure}
	\centering
	\includegraphics[width=1\textwidth]{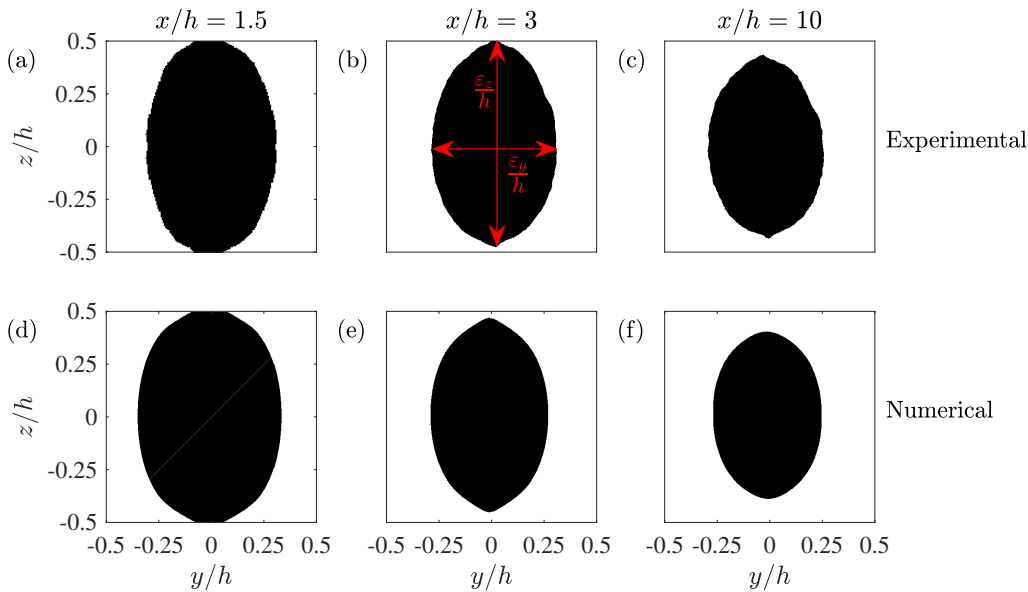} \\
	\caption{Experimental (top) and numerical (bottom) cross sections of the thread. Black colour represents the core fluid while white colour indicates the sheath fluid. The cross-sectional views are taken at $x/h = 1.5$ (panels (a) and (d)), $x/h = 3$ (panels (b) and (e)) and $x/h = 10$ (panels (c) and (f)). 
	}
	\label{fig:fig8_thread_cross_section}
\end{figure}

\begin{figure} 
	\centering
	\includegraphics[width=1\textwidth]{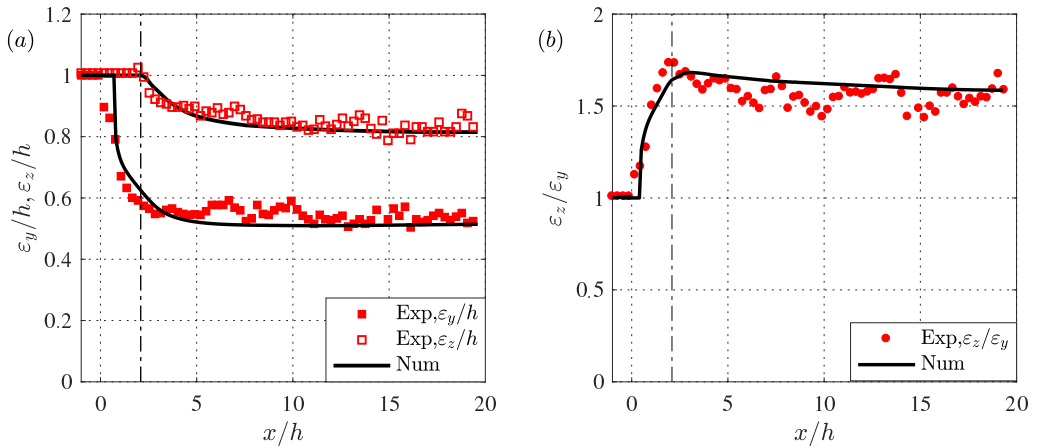} 
	\caption{Thread width $\varepsilon_y/h$ and height $\varepsilon_z/h$ in (a), and aspect ratio  $\varepsilon_z/\varepsilon_y$ (b) as a function of downstream locations $x/h$. The experimental data are shown in red while the numerical data are represented by solid black lines. The vertical dashed-dotted lines at $x/h = 2$ indicate the end of the wetted region.}
	\label{fig:fig9_Aspect_ratio_comparison}
\end{figure}

The cross-sectional views of the thread shape at various downstream positions $x/h=1.5$, $x/h = 3$ and $x/h = 10$ are illustrated in figure~\ref{fig:fig8_thread_cross_section}, and  figure~\ref{fig:fig9_Aspect_ratio_comparison} depicts the length of the ellipsoid axes $\varepsilon_y/h$, $\varepsilon_z/h$ and aspect ratio $\varepsilon_z / \varepsilon_y$ of the cross-section as a function of downstream positions $x/h$. The agreement between the numerical and experimental measurements is superb.  One could also notice that at $x/h = 1.5$, the colloidal dispersion is still being in contact with the top and bottom walls (figure~\ref{fig:fig8_thread_cross_section} (a),(d)) and then after the detachment, the thread becomes thinner and stays nearly stable (figure~\ref{fig:fig8_thread_cross_section} (b),(e) and (c),(f)). This is reflected in the streamwise development of the height~$\varepsilon_z/h$ and the width~$\varepsilon_y/h$ of the thread (see figure~\ref{fig:fig9_Aspect_ratio_comparison}). First, the width decreases much faster than the height and also reaches a constant value at around $x/h\approx5$ whereas there is a very slow change in the height of thread. The values of width and height reached far downstream ($\varepsilon_y/h\approx0.5$ and $\varepsilon_z/h\approx0.8$ at $x/h=20$) are different, confirming that the cross-section of the thread is nearly elliptical as visible in figure~\ref{fig:fig6_Octthread}. These clear differences indicate that the thread formation in the $y$ and $z$ directions could be governed by different physical mechanisms.

\subsection{Shape variations with flow rate ratio\label{flow_rate_ratio}}

\begin{figure} 
	\centering
	\includegraphics[width=1\textwidth]{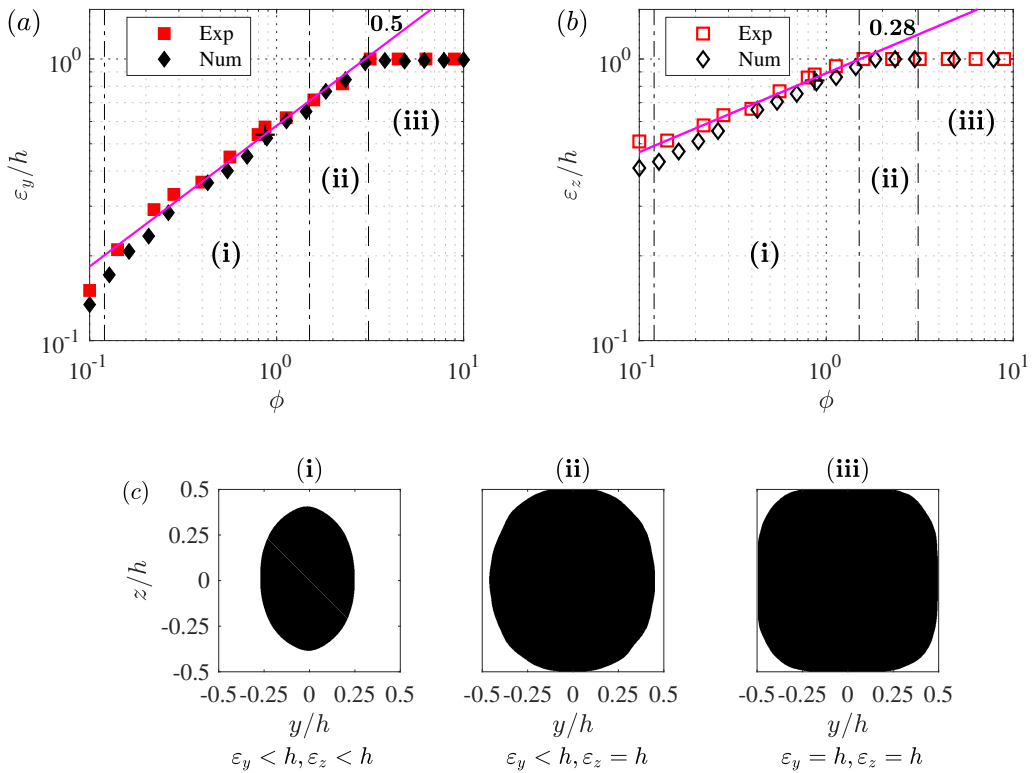}  \\
	\caption{ Width~$\varepsilon_y/h$ and height~$\varepsilon_z/h$ of the thread as a function of flow rate ratio~$\phi$. Red square symbols represent experiments while diamond symbols denote numerical simulations. The fit of the first part of the curves by the relation $\alpha\phi^{k}$ are shown by the pink lines and the fitted values of $k$ are written in each panel. (c) Illustration using numerical data of the different regimes separated in panels (a) and (b) by vertical dashed-dotted lines: threading regime (i) and tubing regime (ii, iii).}
	\label{fig:fig10_Aspect_ratio_fflowrate_comparison}
\end{figure}

In order to investigate the role played by the sheath flows on the thread formation, a parametric variation of the flow-rate ratio $\phi$ over two orders of magnitude has been performed. This is an aspect that is of immense interest when the flow-focusing is used for hydrodynamic assembly of nanofibrils~\citep{haakansson2014hydrodynamic}, since the shape determines the cross-section of an assembled material and could also provide means for synthesizing materials of complex shapes such as rods, discs, ellipsoids~\citep{xu2005}. 
The dimensionless width~$\varepsilon_y/h$ and height~$\varepsilon_z/h$ of the thread measured far downstream  at ~$x/h = 20$ are plotted in log-log scale in figure~\ref{fig:fig10_Aspect_ratio_fflowrate_comparison} as a function of flow-rate ratio~$\phi$. The agreement between experimental and numerical points is very good over the full range of the flow-rate ratio variation. Both curves show two different regimes: for low flow-rate ratio, there is a power law regime $\alpha\phi^{k}$ corresponding to the threading regime, while for high flow-rate ratio, there is saturation at $1$ corresponding to the tubing regime. The fit of the threading regime as a function of flow-rate ratio $\phi$, represented by pink solid lines in figure~\ref{fig:fig10_Aspect_ratio_fflowrate_comparison}, gives $\varepsilon_y/h=0.58\phi^{0.5}$ for the width and $\varepsilon_z/h=0.89\phi^{0.28}$ for the height. A power law in $\phi^{0.5}$ is predicted for a circular thread only controlled by the ratio of the flow rates~\citep{cubaud2009} and has been verified experimentally only by measuring the width of the thread assuming a circular cross section~\citep{Cubaud_2008,CubaudNotaro2014}. The coefficient $0.58$ in front of the power law is also in good agreement with previous studies~\citep{Cubaud_2008,cubaud2009}. This implies that the width of the thread is purely controlled by the flow-rate ratio~$\phi$. However, the height of the thread exhibits a different power law, in~$\phi^{0.28}$. This  indicates that the thread cross section is not circular and that the height of the thread is driven only partly by the flow-rate ratio $\phi$ but could also be dependent on other parameters such as viscosity ratio~$\chi$ and interfacial tension~$\gamma$. Later, in the upcoming Section~\ref{surface_tension}, the effect of EIT~($\gamma$) on the height of thread $\varepsilon_z/h$ is discussed. 

One can also notice that the transitions between the power law regimes happen at different flow-rate ratio values for the width and the height of the thread. There are therefore three different characteristic regimes that can are observed within the range of the flow-rate ratio sampled in this study. They are illustrated in figure~\ref{fig:fig10_Aspect_ratio_fflowrate_comparison} (c). For $\phi<1.5$, the colloidal dispersion forms a thread with an elliptical cross-section (i). In this regime, both width and height of the thread enlarge when increasing $\phi$. In the range $1.5<\phi<3$, the height of the thread saturates and reaches the size of the channel $h$ in $z$ direction while the width keeps increasing. This is the first regime of tubing (ii). Above $\phi>3$, both height and width have saturated and reaches the size of the channel $h$ in the $y$ and $z$ directions. This defines a second tubing regime (iii) where the sheath flows are only localized in the corners of the channel.

\subsection{Effect of interfacial tension\label{surface_tension}}
\begin{figure}
	\centering
	\includegraphics[width=1\linewidth]{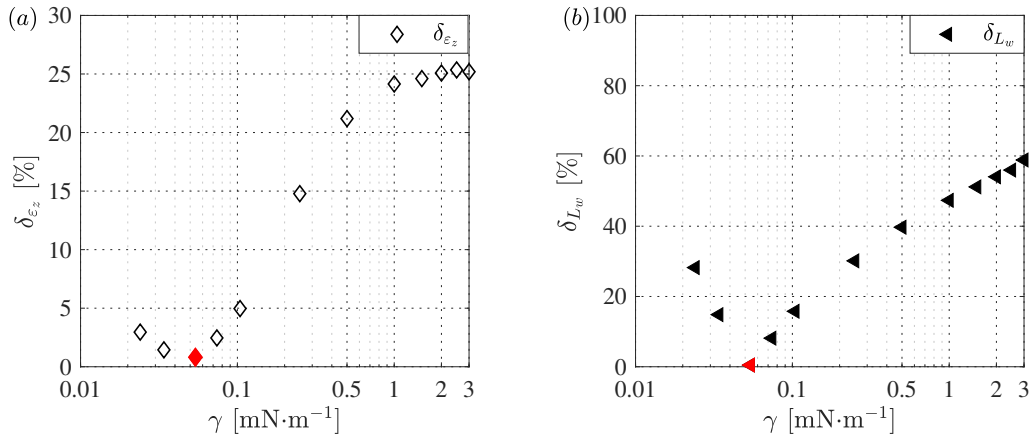} 
	\caption{Variations in percent of the error~$\delta_{\varepsilon_z}$ on the height of the thread~(a) and of the error~$\delta_{L_w}$ on the wetted length~(b) as a function of interfacial tension $\gamma$ obtained from numerical simulations. The red points in panels (a) and (b) indicate the interfacial tension that matches the most the experimental data.}
	\label{fig:fig11_Surface_tension_sweep_diff}
\end{figure}
The transient immiscible behaviour between the two fluids, the colloidal dispersion and its own solvent, provides an opportunity to study the effect and measure the value of so-called effective interfacial tension (EIT)~\citep{truzzolillo2014} for a dynamic interface. Since EIT is a ultra-low and transient phenomena, it is difficult to measure experimentally. We utilize the numerical simulations to quantify values within the range $0.024$~mN$\cdot$m$^{-1}<~\gamma< ~3.000$~mN$\cdot$m$^{-1}$. The height of the thread~$\varepsilon_z/h$ at $x/h=20$ together with the wetted length $L_w/h$ are now used to illustrate how EIT affects the shape in the calculations by comparing with the experimental values. This has been done using the following definitions for the differences,
\begin{equation}
\delta_{\varepsilon_z}(\gamma)=\left|\frac{\varepsilon_z^{num}(\gamma)-\varepsilon_z^{exp}}{\varepsilon_z^{exp}}\right|~~\textrm{and}~~\delta_{L_w}(\gamma)=\left|\frac{L_w^{num}(\gamma)-L_w^{exp}}{L_w^{exp}}\right|.
\end{equation}
The differences $\delta_{\varepsilon_z}(\gamma)$ and  $\delta_{L_w}(\gamma)$ are plotted in figure~\ref{fig:fig11_Surface_tension_sweep_diff}. Both curves show significant variations i.e., $\varepsilon_{z}/h$ and $L_w/h $  are controlled by $\gamma$ for a fixed viscosity ratio~$\chi$ and flow-rate~ratio~$\phi$. The minima of  $\delta_{\varepsilon_z}(\gamma)$ and  $\delta_{L_w}(\gamma)$, are indicated by red filled symbols in figures~\ref{fig:fig11_Surface_tension_sweep_diff}~(a) and (b) and occur at $\gamma  = 0.054$~mN$\cdot$m$^{-1}$. This value is used for the comparison of numerical results with the experimental measurements in this study. Moreover, the order of magnitude of $\gamma \sim 10^{-2}$~mN$\cdot$m$^{-1}$,  is in line with the recent experimental observations of EIT made by~\cite{truzzolillo2016} for miscible complex fluids like colloidal and polymer dispersions where there is a presence of dynamic interface due to weak diffusion. Thus, the presence of EIT in this study is a consequence of the high Péclet number leading to the existence of a \textit{de facto} dynamic interface.

\begin{figure}
	\centering
	\includegraphics[width=1\linewidth]{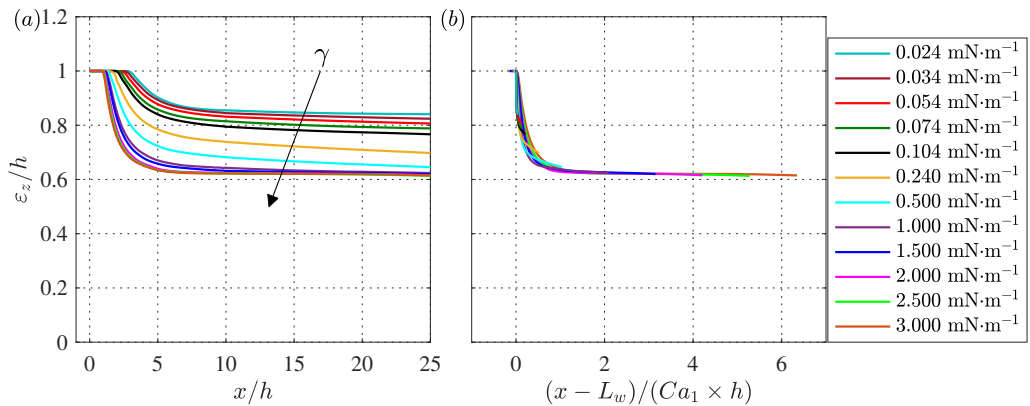} 
	\caption{Variations in the height of the thread~${\varepsilon_z/h}$ ~(a) with downstream locations~$x/h$ for various effective interfacial tension $\gamma$~(b) as a function of scaled downstream location for different effective interfacial tension $\gamma$. $Ca_{1}$ corresponds to the effective capillary number of core fluid thread.} 
	\label{fig:fig12_elliptical_circular}
\end{figure}

Figure~\ref{fig:fig12_elliptical_circular}~(a) shows the height of the thread~$\varepsilon_z/h$ as a function of downstream position~$x/h$ for various $\gamma$. First, one can note that the wetted length $L_w/h$ is slightly influenced by $\gamma$, as the detachment of the various threads from the top and bottom walls of the channel takes place in the range $x/h=1$ - $3$. Once the thread is detached, the larger the $\gamma$, the faster the thread converges towards a value of $0.63$, corresponding to the width of the thread $\varepsilon_y/h$ for the simulation with the highest $\gamma$. Far downstream, for $x/h>20$, the height of the threads with low $\gamma$ is still decreasing. This strongly suggests that the near-elliptical shape of the thread cross-section is only a transient state, before flattering towards circular shape. Indeed, the thread will eventually become circular as the effective interfacial tension tends to minimize the contact surface. However, the time scale for the thread to converge towards a circular cross-section appears to be dependent on the EIT~$\gamma$.

For instance, let us consider the two-dimensional problem of the thread cross-section, in the $y$-$z$ plane as shown in Figure~\ref{fig:fig8_thread_cross_section}. Just after the detachment, at $x=L_w$, the cross-section has an near-elliptical shape, with $\varepsilon_z/h\approx1$ and $\varepsilon_y/h<1$. The thread reorganizes itself by converging towards a circular cross-section, with $\varepsilon_z/h=\varepsilon_y/h<1$. This dynamics is driven by ~$\gamma$ at the interface. Indeed, as the interface between the core and sheath fluids is curved, there is a Laplace pressure difference between the two fluids, proportional to the $\gamma$ and to the local curvature of the interface. As $\varepsilon_z/h>\varepsilon_y/h$ and assuming a constant pressure in the sheath flow, the pressure on the top and bottom of the thread is thus larger than the pressure on the sides. This in turn leads to a pressure gradient $\delta P$ within the thread, proportional to EIT ($\gamma$) and dependent on the geometry of the thread, driving the thread cross-section from an near-elliptical to a circular shape.

As the viscosity ratio $\chi$ in the system is very high, one can neglect the influence of the sheath flows on this dynamics. Consequently, it is mainly dependent on the viscosity $\eta_1$ of the core fluid. The typical time scale $\tau$ for this dynamics can be estimated \textit{via} the Stokes equation as
\begin{equation}
\tau=\frac{\eta_1}{\delta P}\propto\frac{\eta_1}{\gamma},
\end{equation}
where the thread geometry dependence of $\delta P$ is hidden in the proportional symbol. During this time scale, the thread cross-section is advected downstream with a velocity~$U$. As the flow rate is conserved, this velocity is dependent on the shape of the thread $U=4Q_1/(\pi\varepsilon_z\varepsilon_y)$, but we neglect these variations for simplification and consider that $U \approx Q_1/h^2$. Accordingly, the typical distance $x_c$ needed for the thread to converge from near-elliptical to circular cross-section can be estimated by
\begin{equation}
x_c-L_w=U \tau \propto \frac{\eta_1 Q_1}{\gamma h^2}.
\end{equation}
Similar to immiscible fluids, one can therefore define an effective capillary number for the thread, $Ca_1=\eta_1Q_1/(h^2\gamma)$, and normalize the modified downstream location $(x-L_w)/h$ with it. This is shown in Figure~\ref{fig:fig12_elliptical_circular}~(b). All the  numerical data collapse well on a master curve, showing that the dynamics of the height of the thread is governed by this effective capillary number. To measure the EIT acting between two miscible fluids with high Péclet and high viscosity ratio, it is therefore only necessary to measure the evolution of the height of the thread $\varepsilon_z/h$ formed by these two fluids and to compare it with the master curve as depicted in Figure~\ref{fig:fig12_elliptical_circular}~(b).

It is worthwhile to note that \cite{cubaud2009} made a circular prediction for a square-cross section channel of width $h$ assuming a circular annular flow, which means that the way to pour liquids in the channel was not taken into account. In a flow-focusing geometry, there is no rotational symmetry of $\frac{\pi}{2}$ as for a square duct. As a consequence, the thread only respects the two planes of symmetry of the channel at $y/h=0$ and $z/h=0$. The thread therefore detaches from the channel wall with an elliptical cross-section, thinner in the direction of the sheath inlets, i.e. the $y$ direction because the sheath has higher momentum in that direction.

\section{Velocity fields\label{velocity_fields}}

\begin{figure}
	\centering
	\includegraphics[width=0.8\textwidth]{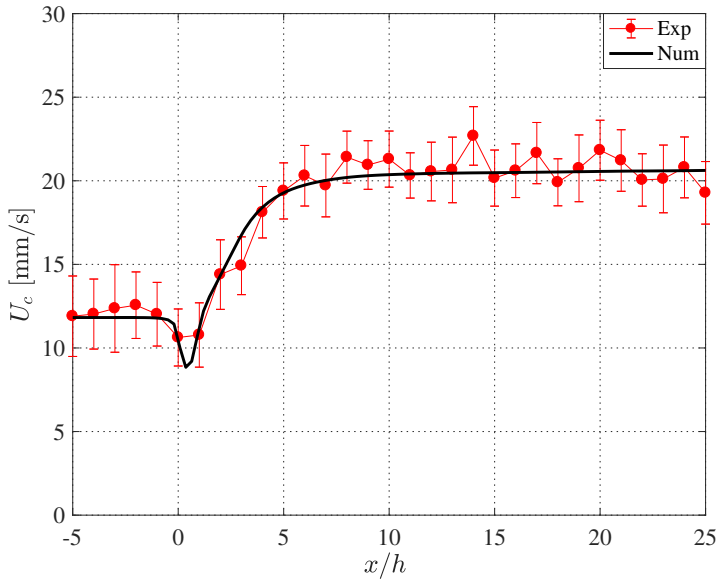}
	\caption{Experimental (red points) and numerical (solid black line) centreline velocity~$U_c$ as a function of the downstream position~$x/h$. Error bars for experimental data are shown and are around $2\%$.}
	\label{fig:fig12_centrelinevelocity}
\end{figure}

In addition to the comparison between numerical simulations and experiments with respect to shape and topology of the thread, we will also compare the velocity fields. These are responsible for the alignment of the nanofibrils in the dispersion~\citep{jeffery1922} and are therefore of importance to material fabrication. 

In figure~\ref{fig:fig12_centrelinevelocity}, the centerline velocity $U_c$ is shown as a function of the downstream location $x/h$, obtained by both experimental measurements and numerical simulations. The velocity is first constant before the focusing region ($x/h<0$) and right after the injection of the sheath flows at $x/h = 0$, there is a minor deceleration followed by rapid acceleration, before almost stabilizing at around $x/h=10$. The increasing velocity implies that there is an acceleration of the core flow, along the direction ($x$-direction) of the channel. Thus the flow-focusing geometry through sheath flows creates an extensional flow with a velocity gradient in the direction of the core flow, leading to the alignment of the nanofibrils in the colloidal dispersion~\citep{H_kansson_2016}.

\begin{figure}
	\centering
	\includegraphics[width=1\textwidth]{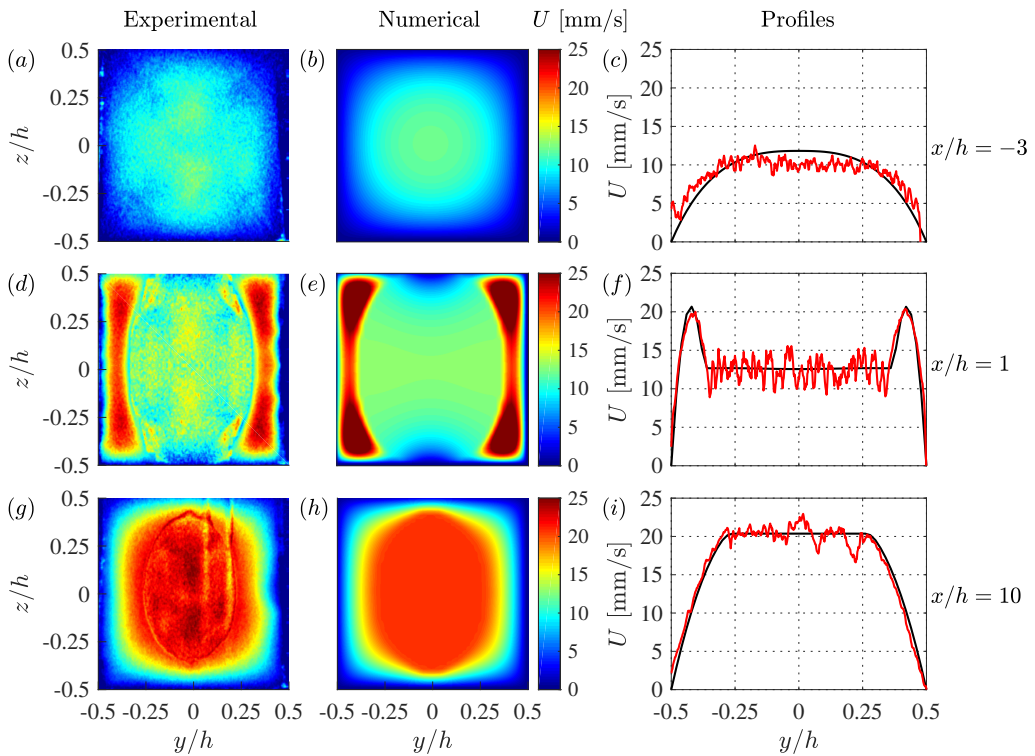}
	\caption{
		Experimental and numerical velocity maps (first and second column) and profiles (third column) at three different downstream locations: $x/h=-3$ (top row), $x/h=1$ (center row) or $x/h=10$ (bottom row). The velocity profiles in panels (c), (f) and (i) show the $x$-component of the velocity at $z/h=0$. The experimental data are represented by the red line while the numerical data are shown with the black line.}
	\label{fig:fig13_cross_sectionalvelocity}
\end{figure}

Figure~\ref{fig:fig13_cross_sectionalvelocity} shows experimental and numerical maps and profiles of the $x$-component of the velocity in different cross-sectional $x/h$ planes: before the focusing region~($x/h=~-3$), just after the focusing region ($x/h=1$) and at far downstream ($x/h=10$). Both the velocity maps and profiles show again an excellent quantitative agreement. 
Before the focusing point, the velocity profile is near parabolic. Just after the focusing, at the beginning of the acceleration, one can see that the colloidal dispersion is still attached to the walls at $z/h=\pm0.5$, in the wetted region, but is surrounded by the sheath flows in the $y$-direction. The sheath flows are visible on the velocity maps as the region with the highest velocities on the sides. The thread exhibits a uniform velocity profile in the centre. Far downstream, the thread has attained its nearly elliptical shape and steady velocity flow-field. The velocity profile within the thread is completely uniform indicating plug flow. Surrounding the colloidal dispersion, the sheath flow velocity profiles are parabolic, connecting the thread to the walls of the channel. This has been predicted by~\cite{cubaud2009} for a circular thread in a square channel.

\section{Conclusions\label{conclusion}}

In this paper, we present a comprehensive comparison of experimental and numerical results on 3D thread formation under the influence of EIT with a miscible fluid pair of colloidal dispersion and its own solvent as a model system. With a  high Péclet number, this system is a weakly diffusive \textit{i.e.}, the timescale for diffusion of the nanofibrils from the core flow into the sheath flows being substantially larger than the residence time of the two fluids in the channel. Therefore, the system exhibits a near-immiscible behaviour with a \textit{de facto} dynamic interface where EIT is acting. Three-dimensional versions of the finite volume based open-source code OpenFOAM~\citep{OpenCFD} with volume of~fluid~(VOF) method have been used for the numerical simulations and experimental measurements were carried out with Optical Coherence Tomography (OCT). The validation of the numerical scheme was performed by choosing and reproducing a reference case for immiscible fluids reported in~\cite{Cubaud_2008} that demonstrates the rich dynamics of the flow features typically seen in a microfluidic flow-focusing configuration. Our simulations capture all the features of various flow patterns: threading, jetting, dripping and tubing regimes, in very good agreement with the experiments of \cite{Cubaud_2008} both in terms of flow visualisations and in the state space mapped based on capillary numbers of two fluids.

In the threading regime, the 3D numerical simulations of the present study correctly reproduce the experimental observations in terms of thread topology and morphology of the wetted region. The measurements and simulations of the cross-sectional width $\varepsilon_y/h$ and height $\varepsilon_z/h$ show the same behaviour where the thread width reaches a constant value faster than the height, and where the cross-sectional shape of the thread far downstream is nearly elliptic. The shape variation of width $\varepsilon_y/h$ and height $\varepsilon_z/h$ of the thread as a function of flow-rate ratio $\phi$ show good agreement between experiments and numerical simulations revealing interesting features: the width and height of thread both exhibit power-law  behaviour with flow-rate ratio $\phi$. The width of the thread scales purely with the flow-rate ratio as $\phi^{0.5}$ in good agreement with previous studies~\citep{cubaud2009, Cubaud_2008}. However, the height of thread scales as $\phi^{0.28}$ confirming the nearly elliptical cross-sectional shape, thus indicating the dependence of height on other parameters like EIT along with the flow-rate ratio $\phi$.

The three-dimensional numerical simulations gave access to probe for a detailed investigation of the influence of EIT by comparing with the elliptical thread shape given by experiments. Thus, it provided a framework to deduce the values of EIT by minimizing the differences between simulations and experiments. Effective interfacial tension was determined by varying the surface tension in the simulations and comparing the numerically generated data with the experimental wetted length~$L_w/h$ and height~$\varepsilon_z/h$ of the thread.  The  minimum differences are found to be very small, typically about less than~1\%. Furthermore, by estimating the time and length scales, we show that the transition of the thread shape from elliptical to circular is governed by so called effective capillary number. This proves the significance of EIT on the thread dynamics in the microfluidic flow system. 

Also, we show that the numerically simulated velocity flow fields along the centreline and across the cross-section of the channel, show an excellent agreement with the experimental measurements obtained through Doppler OCT. The transversal profile of the streamwise velocity  divulges the parabolic nature of core flow before the focusing. Once the dispersion thread detaches from the walls and attains a near-elliptical shape, its velocity profile is flat. The water sheath flow close to the walls has a parabolic profile. 

Summing up, our three-dimensional experiments and numerical simulations in a microfluidic system have shed new light on the significance and impact of EIT present in miscible complex fluids such as colloidal and polymer fluids, unveiling a distinct framework for assembly, transport and control of material properties through the optimal usage of the surface area of thread shape and wetted region along with other flow parameters such as flow-rate and viscosity ratios. Two-dimensional observations of microfluidic experiments cannot capture this fascinating flow physics. Furthermore, the developed methodology combining experiments and numerical simulations can be used to measure effective interfacial tension acting between two miscible fluids. 

\section*{Acknowledgements}

Financial support by the Swedish Research Council for Environment,
Agricultural Sciences and Spatial Planning (FORMAS), \& Knut and Alice Wallenberg foundation through WWSC is gratefully acknowledged. Computer time  was  provided  by  the  Swedish  National Infrastructure for Computing (SNIC). 
The authors thank Dr. Mathias Kvick, Calvin Brett and Dr. Jingmei Li for experimental assistance.

\appendix 
\section{\\Optical coherence tomography measurements and~analysis \label{OCT}}
\label{sec:Appendix}

This appendix gives some details about the data acquisition and processing using OCT. The simplest acquisition is a scan of the sample along the depth, i.e. the direction of the light in the sample (vertical in figure~\ref{fig:fig3-working-principle-OCT}). Such a scan is called an A-scan and the frequencies given below corresponds to the acquisition frequencies of this type of scans. To obtain a two-dimensional (B-scan) or a three-dimensional image of the sample, it is necessary to perform multiple A-scans by moving laterally the beam in either one or two orthogonal directions. Three-dimensional scans have been used to measure the shape of the thread while the velocity acquisition is only limited to two-dimensional scans. 

The OCT apparatus is placed vertically and the channel is inclined with an angle of~$7^{\circ}$ with respect to the horizontal in order to limit the apparition of multiple reflections of light rays created by the sample and reverberated back to OCT. As the colloidal dispersion and water exhibit differences in optical scattering, the different fluid phases are separated after image binarization using a contrast threshold. The frequency of acquisition is $f=5.5$~kHz for 3D acquisitions to determine the shape of the thread, leading to a total scanning time of a few minutes. The scanning time is longer with respect to the typical convective time scale $T_c$ but it is acceptable since the system is in a stationary state. The Doppler acquisitions have been done at $f=5.5$~kHz or at $28$~kHz, depending on the maximum velocity in the measured region of the thread. Typical scanning time is less than a second and the velocity measurements have been averaged over $100$ repetitions.

In order to measure the velocity fields using Doppler velocimetry, it is necessary to seed the fluids with scattering particles. For the water sheath flows, this is done by adding commercial milk to reach $10\%$ concentration, only when measuring velocity profiles. The properties of the sheath flows are not changed by this small modification. For the colloidal dispersion, light scattering already exists without any addition of particles but is relatively low. However, we have chosen not to add particles in the dispersion in order to preserve its physiochemical properties. This low scattering in the dispersion leads to incorrect estimations of the absolute velocity magnitude but to correct relative velocity variations, i.e. the measured velocity profiles are fully valid but need to be rescaled. This rescaling is done by integrating the measured velocity fields $U_{meas}(y,z)$ in the thread and normalizing them by the flow rate $Q_1$ prescribed by the syringe-pump:
\begin{equation}
U_{corr}(y,z)=U_{meas}(y,z) \frac{Q_1}{\iint_{\Sigma} U_{meas}(y,z) \textrm{d}y\textrm{d}z},
\end{equation}
where $U_{corr}(y,z)$ are the corrected velocity fields and $\Sigma$ represents the thread cross-section.

\bibliographystyle{jfm}
\bibliography{jfm-references.bib}

\end{document}